\documentclass{article}
\usepackage{spconf,amsmath,graphicx}
\usepackage[utf8]{inputenc}
\usepackage[table]{xcolor}
\usepackage{color}
\usepackage{multirow}
\usepackage{amsfonts,amssymb}
\usepackage{float}
\usepackage[ruled]{algorithm2e}

\definecolor{gris}{cmyk}{0.02, 0.02, 0.01, 0}

\title{LIA system description for NIST SRE 2016}
\name{Mickael Rouvier, Pierre-Michel Bousquet, Moez Ajili, Waad Ben Kheder}
\secondlinename{Driss Matrouf, Jean-Fran\c cois Bonastre}

\address{LIA, Universit\'e d'Avignon, France}
\begin{document}
%
\maketitle
\begin{abstract}
This paper describes the LIA speaker recognition system developed for the Speaker Recognition Evaluation (SRE) campaign. Eight sub-systems are developed, all based on a state-of-the-art approach: i-vector/PLDA which represents the mainstream technique in text-independent speaker recognition. These sub-systems differ: on the acoustic feature extraction front-end (MFCC, PLP), at the i-vector extraction stage (UBM, DNN or two-feats posteriors) and finally on the data-shifting (IDVC, mean-shifting). The submitted system is a fusion at the score-level of these eight sub-systems.
\end{abstract}

\section{Introduction}
This paper describes the systems developed by the LIA for the 2016 National Institute of Standard and Technology (NIST) Speaker Recognition Evaluation (SRE). 

LIA developed eight sub-systems which are described in this paper. The eight sub-systems are based on i-vector/PLDA paradigm. I-vector/PLDA paradigm is the state-of-the-art approach in speaker recognition. The i-vector approach provides an elegant way of reducing a large-dimensional input vector (representing the speaker data) to a small-dimensional feature vector~\cite{Dehak11_a}. I-vectors are extracted on total variability space, no distinction is made between speaker and channel variation. Probabilistic Linear Discriminant Analysis (PLDA) is used to disentangle speaker effects from other sources of undesired variability~\cite{Ioffe06,prince2007probabilistic}.

The sub-systems are constructed by combining different front-end and back-end. The sub-systems differ from acoustic features (MFCC or PLP), i-vector extraction (UBM, DNN or two-feats posteriors) and data-shifting (IDVC or mean-shifting). The submitted system is a fusion at the score-level of these eight sub-systems. All the sub-systems are mainly based on the open-source ALIZE toolkit, freely available~\footnote{http://alize.univ-avignon.fr}.

This paper is organized as follows: In Section~\ref{sec:data}, we present the different datasets used in our systems. We briefly describe our system in Section~\ref{sec:overview}. Details of the submitted systems and the components are described in Section~\ref{sec:component}. The results for the individual sub-systems and the fused system are presented in Section~\ref{sec:results}. Finally in Section~\ref{sec:computationtime} we show the CPU time and memory requirements for computing the score of one verification trial.

\section{Train and development datasets}
\label{sec:data}

The Table~\ref{table:datasets} lists the different datasets that we used for training the system.

\begin{table}[htb]
\centering
\scalebox{0.8}{
\begin{tabular}{|l|l|}
\hline
Dataset(s) & Task \\
\hline
\hline
LDC2016E46\_SRE16\_Cal\_My\_Net & Development Set \\
\hline
LDC2016E46 (unlabeled) & i-vector normalization \\
\hline
SRE’04, 05, 06, 08 & \multirow{5}{*}{UBM, T matrix, PLDA, IDVC} \\
Switchboard-2 Phase II & \\
Switchboard-2 Phase III & \\
Switchboard Cellular Part 1 & \\
Switchboard Cellular Part 2 & \\
\hline
Switchboard-1 Release 2 & DNN \\
\hline
\end{tabular}
}
\caption{\label{table:datasets}Datasets used for training the system.}
\end{table}

\section{Overview of the system}
\label{sec:overview}



The different steps of our i-vector systems can be summarized as follows:

\begin{itemize}
\item \textbf{Feature extraction} : Two acoustics features are extracted : Mel-Frequency Ceptral Coefficients (MFCC) and Perceptual Linear Prediction (PLP).

\item \textbf{Voice activity detection (VAD)} : remove silence and low energy speech based on the C0 component.

\item \textbf{I-vector extraction} : three different kind of i-vectors are extracted (GMM/i-vector, DNN/i-vector and two-feats/i-vector). These i-vectors differ from the manner in which the statistics are collected.


\item \textbf{Pre-normalization} : The raw i-vectors extracted are first whitened  and length-normalized ($\mathbf{LW}$-normalization).

\item \textbf{Data-shifting} : considering the language mismatch between training and development corpus. Two data-shifting methods are used in order to compensate this mismatch : Inter Dataset Variability Compensation (IDVC) and mean-shifting.


\item \textbf{PLDA learning} : a non-classical PLDA is learned on the i-vectors. This PLDA is fully described in Section~\ref{sec:PLDA}.


\item \textbf{Post-normalization} : The between- and within-class covariance matrix etimated by the PLDA are diagonalized and a new $\mathbf{LW}$-normalization is applied on the i-vectors.


\item \textbf{PLDA scoring} : a verification score is calculated.


\end{itemize}

\section{System components}
\label{sec:component}

\subsection{Acoustic features}

Before feature extraction, all waveforms are first down-sampled to 8 kHz, and blocked into 25 ms frames with a 10 ms skip-rate. Two acoustic features are extracted using Kaldi toolkit~\cite{Povey_ASRU2011} : MFCC and PLP. All features use 20 cepstral coefficients and log-energy, appended with the first and second order time derivatives, thus providing 60 dimensional feature vectors. A cepstral mean normalization is applied with a window size of 3 seconds.

\subsection{Voice Activity Detection}

VAD removes silence and low energy speech segments. A simple energy-based VAD is used based on the C0 component of the acoustic feature. The algorithm is based on thresholding the log-energy and taking the consensus of threshold decisions within a window of 11 frames centred on the current frame.

\subsection{i-vector}

An i-vector extractor is a data-driven front-end that maps temporal sequences of feature vectors (e.g., MFCC or PLP) into a single point in a low-dimensional vector space. This is accomplished by collecting sufficient statistics. In these evaluation the statistics are obtained from : Universal Background Model (UBM), Deep Neural Network (DNN) and two-feats. All the extracted i-vectors are 400-dimension.


\subsubsection{UBM}
The generation of i-vectors requires use of UBM which modelizes the generation of the acoustic features (cepstrum + first and second derivatives). The UBM used here is a GMM (Gaussian Mixture Model) of 4096 Gaussians, where each Gaussian is characterized by its mean and its full-covariance matrix. The UBM is trained on SRE'04-08 and Switchboard corpus, using the standard EM algorithm. 
\subsubsection{DNN}

In~\cite{lei2014novel}, authors propose to collect statistic by using a DNN that are trained to classify phoneme states. DNN is trained with 4 hidden layers. The input layer takes 60 dimensional MFCC features with 7-frame temporal context and cepstral mean subtraction (CMS) performed over a window of 6 seconds. Each hidden layer has 1024 nodes. The ouput dimension is 4096 senone. The forced alignment between the state-level transcripts and the corresponding speech signals by the GMM/HMM triphon system is used to generate labels for DNN training.

\begin{table*}
\centering
\caption{Details of the subsets used for IDVC method}
\label{table:Table_IDVC_subsets}
\begin{tabular}
[c]{|c|l|l|l|l|l|}\hline
subset & language & native language & gender & \#segments & \#speakers\\\hline\hline
1 & english & english & F & 13934 & 774\\\hline
2 & english & english & M & 10379 & 504\\\hline
3 & english & non english & F & 6087 & 784\\\hline
4 & english & non english & M & 9326 & 517\\\hline
5 & non  english & all & F & 4576 & 663\\\hline
6 & non  english & all & M & 2868 & 413\\\hline
additional & --- & --- & all & 2272 & ---\\\hline
\end{tabular}
\end{table*}

\subsubsection{two-feats}

It is well known that the sucess key of the i-vector paradigm is the robustness of the a-posteriori probabilities estimation against the UBM. To increase the robustness of this estimation, we propose to use two acoustic features streams instead of only one: PLP based stream and MFCC based stream. To do so, we firstly estimate a PLP based UBM and for each frame, we generate the a posteriori probabilities. Then, we use the MFCC frames and these last a posteriori probabilities to estimate the parameters of the MCCF based UBM (one EM iterartion). At the end of this process, we obtain two UBMs having the same topology: same number of Gaussians with correspondance between pairs of Gaussians having the same index in the two UBMs (PLP and MFCC). From now on, the a posteriori probablities for a given frame is obtained by combining the ones coming from the PLP-UBM and MFCC-UBM. Lets, $P_{mfcc}=[p_{1}^ {mfcc},...,p_{4096}^{mfcc}]$ and   $P_{plp}=[p_{1}^ {plp},...,p_{4096}^{plp}]$, the final a posteriori porobabilities are given by:
$$
p_i=\frac{p_i^{plp}*p_i^{mfcc}}{\sum_{j}{p_j^{plp}*p_j^{mfcc}}}
$$


\subsection{Pre-normalization}

I-vectors are whitened and length-normalized in order to make them more Gaussian, and also to reduce the shift between training and test data, as remarked in \cite{Cumani2016}. The whitening technique we use for NIST SRE 2016 evaluation is a standardization according to the within-class covariance matrix $\mathbf{W}$, as proposed in \cite{Dehak11_a,Bousquet2012}. We denote by $\mathbf{LW}$ this transformation (Length-normalization of $\mathbf{W}$-standardized vectors).

\subsection{Data-shifting}

\subsubsection{Inter dataset variability compensation}

In order to reduce the shifts of language and gender, we include in our system the Inter Dataset Variability Compensation (IDVC) technique, as described in \cite{Aronowitz2014}. This technique seeks to compensate eventual mismatches, between training and test data, by removing unexpected variability of model parameters. We apply this technique to limit the uncertainty of mean and within-speaker covariance matrix involved by gender and language mismatches. The IDVC method is trained on 6 subsets of segments from NIST\ SRE\ 2004, 2005, 2006, 2008 and the additional subset of development data from the evaluation major language provided by NIST (the latter only for mean-subspace removal, as this subset is unlabeled). Table~\ref{table:Table_IDVC_subsets} details the content of these subsets.

\subsubsection{Mean-shifting}

Mean-shifting calculates the mean of the Call-My-Net development data and subtract it to the test i-vectors.

\subsection{PLDA}
\label{sec:PLDA}

\subsubsection{Learning}

LIA systems use the PLDA learning proposed in the Kaldi toolkit
\cite{Kaldi, Povey_ASRU2011}. Given a set of $n_{s}$ vectors from a
training speaker $s$, this model assumes that the centered mean vector $m_{s}$
of speaker $s$ can be decomposed as:%

\[
m_{s}=x_{s}+y_{s}%
\]

where%

\begin{align}
x_{s} &  \sim\mathcal{N}\left(  0,\mathbf{B}\right)  \nonumber\\
y_{s} &  \sim\mathcal{N}\left(  0,\frac{1}{n_{s}}\mathbf{W}\right)
\end{align}

$\mathbf{B}$ (resp. $\mathbf{W}$) denoting the between (resp. within)-class
covariance matrix and $\mathcal{N}\left(  .\right)  $ the Gaussian pdf. Thus,
this model takes into account some uncertainty about the speaker mean
position, depending on the size $n_{s}$ of its training set.

Starting from deterministic estimations of $\mathbf{B}$ and $\mathbf{W}$, an
EM-like iterative algorithm is applied to optimize these matrices. It can be
shown that the distributions of $x_{s}$ and $y_{s}$ a posteriori of $m_{s}$
are Gaussian:%

\begin{align}
x_{s}|m_{s}  & \sim\mathcal{N}\left(  w_{s},\mathbf{M}_{s}\right)  \\
y_{s}|m_{s}  & \sim\mathcal{N}\left(  m_{s}-w_{s},\mathbf{M}_{s}\right)
\end{align}

where%

\begin{align}
\mathbf{M}_{s} &  =\left(  \mathbf{B}^{-1}+n_{s}\mathbf{W}^{-1}\right)
^{-1}\nonumber\\
w_{s} &  =n_{s}\mathbf{M}\mathbf{W}^{-1}m_{s}%
\end{align}

It can also be shown that the contributions of this speaker-class to the
between and within-class covariance are respectively equal to:%

\begin{align}
E\left[  x_{s}x_{s}^{t}\right]   &  =w_{s}w_{s}^{t}+\mathbf{M}_{s}\\
E\left[  y_{s}y_{s}^{t}\right]   &  =\left(  m_{s}-w_{s}\right)  \left(
m_{s}-w_{s}\right)  ^{t}+\mathbf{M}_{s}%
\end{align}

Thus, $\mathbf{B}$ and $\mathbf{W}$ can be updated as follows:%

\begin{align*}
\mathcal{B} &  =\frac{1}{S}\sum\nolimits_{s}\left(  w_{s}w_{s}^{t}%
+\mathbf{M}_{s}\right)  \\
\mathcal{W} &  =\frac{1}{N}\sum\nolimits_{s}n_{s}\left(  \left(  m_{k}%
-w_{k}\right)  \left(  m_{k}-w_{k}\right)  ^{t}+\mathbf{M}_{k}\right)
\end{align*}

where $S$ is the number of training speakers and $N=\sum_{s}n_{s}$ is the total 
amount of
observations. The $\left(  \mathbf{B},\mathbf{W}\right)$ learning algorithm is 
described below:

\subsubsection{Algorithm}%

\begin{tabular}
[c]{|l|l|l}%
\multicolumn{3}{|l}{$S=$ number of classes; $n_{s}=$ number of observations
for speaker $s$}\\
\multicolumn{3}{|l}{$N=\sum\nolimits_{s}n_{s}$ total amount of observations}\\
\multicolumn{3}{|l}{Compute initial $\mathbf{B}$ and $\mathbf{W}$}\\
\multicolumn{3}{|l}{$m_{all}=\frac{1}{S}\sum\nolimits_{s}m_{s}=\frac{1}{S}%
\sum\nolimits_{s}\left(  \frac{1}{n_{s}}\sum\nolimits_{x\in s}\mathbf{w}\right)
$}\\
\multicolumn{3}{|l}{for $iter=1$ to $nb\_iterations$}\\
& \multicolumn{2}{|l}{$\mathcal{B}=0$; $\mathcal{W}=0$}\\
& \multicolumn{2}{|l}{for each speaker $s$}\\
&  & $m_{s}$ $\leftarrow m_{s}-m_{all}$ (centered mean)\\
&  & $\mathbf{M}_{s}=\left(  \mathbf{B}^{-1}+n_{s}\mathbf{W}^{-1}\right)
^{-1}$\\
&  & $w_{s}=n_{s}\mathbf{M}_{s}\mathbf{W}^{-1}m_{s}$\\
&  & $\mathcal{B}\leftarrow\mathcal{B}+w_{s}w_{s}^{t}+\mathbf{M}_{s}$\\
&  & $\mathcal{W}\leftarrow\mathcal{W}+n_{s}\left(  m_{s}-w_{s}\right)
\left(  m_{s}-w_{s}\right)  ^{t}+n_{s}\mathbf{M}_{s}$\\
& \multicolumn{2}{|l}{$\mathbf{B}\leftarrow\mathcal{B}/S$ ; $\mathbf{W}%
\leftarrow\mathcal{W}/N$}%
\end{tabular}

\begin{table*}
\setlength{\tabcolsep}{2pt} 
\renewcommand{\arraystretch}{1} 
\centering
\begin{tabular}{|l|*{2}c|*{2}c|*{2}c|*{2}c|*{2}c|*{2}c|}\hline
 \multirow{3}{*}{Feat/I-vec/Shifting} & \multicolumn{4}{c|}{\bf Mandarin (CMN)} & 
\multicolumn{4}{c|}{\bf Cebuano (CEB)} & \multicolumn{2}{c|}{\bf Equalized} &
\multicolumn{2}{c|}{\bf Unequalized} \\ 
\cline{2-13}
  &\multicolumn{2}{c|}{Male} & \multicolumn{2}{c|}{Female} & 
\multicolumn{2}{c|}{Male} & \multicolumn{2}{c|}{Female} & 
\multicolumn{2}{c|}{Male+Female} &
\multicolumn{2}{c|}{Male+Female} \\
\cline{2-13}
 & EER(\%) & minC &  EER(\%) & minC  &  EER(\%) & minC  & 
 EER(\%) & minC &  EER(\%) & minC  &  EER(\%) & minC \\ \hline\hline
1 - MFCC/UBM/IDVC &
6.19 & 0.323 & 14.56 & 0.644 & 22.92 & 0.820 & 23.16 & 0.840 & 16.71 & 0.657 & 16.67 & 0.721 \\ 
\hline
2 - MFCC/UBM/Mean &
7.76 & 0.34 & 18.64 & 0.689 & 25.08 & 0.849 & 22.85 & 0.812 & 18.58 & 0.672 & 19.03 & 0.699 \\ 
\hline
3 - PLP/UBM/IDVC &
5.04 & 0.223 & 15.49 & 0.654 & 23.17 & 0.855 & 25.19 & 0.881 & 17.22 & 0.65 & 17.36 & 0.779
\\ \hline
4 - PLP/UBM/Mean &
7.07 & 0.282 & 16.79 & 0.657 & 24.10 & 0.892 & 23.29 & 0.803 & 17.81 & 0.658 & 18.33 & 0.732 \\ 
\hline
5 - MFCC/DNN/IDVC &
5.47 & 0.292 & 16.75 & 0.710 & 27.50 & 0.937 & 25.71 & 0.878 & 18.86 & 0.704 & 18.58 & 0.783 \\ 
\hline
6 - MFCC/DNN/Mean & 
6.75 & 0.310 & 20.88 & 0.727 & 27.56 & 0.960 & 25.17 & 0.865 & 20.09 & 0.716 & 20.97 & 0.768 \\ 
\hline
7 - MFCC/two-feats/IDVC &
6.15 & 0.267 & 15.32 & 0.607 & 24.08 & 0.844 & 23.20 & 0.832 & 17.19 & 0.638 & 17.44 & 0.729 \\ 
\hline
8 - MFCC/two-feats/Mean & 
7.56 & 0.296 & 18.60 & 0.641 & 25.58 & 0.858 & 23.49 & 0.784 & 18.81 & 0.644 & 19.02 & 0.687 \\ 
\hline
\end{tabular}
\caption{\label{table:performance}Performance of our 8 single 
systems: MFCC/PLP/DNN/two-feats with two techniques of data-shifting: IDVC or mean-shifting (substraction of mean).}
\end{table*}

\begin{table*}
\setlength{\tabcolsep}{2pt} 
\renewcommand{\arraystretch}{1} 
\centering
\begin{tabular}{|l|*{2}c|*{2}c|*{2}c|*{2}c|*{2}c|*{2}c|}\hline
 \multirow{3}{*}{System} & \multicolumn{4}{c|}{\bf Mandarin (CMN)} & 
\multicolumn{4}{c|}{\bf Cebuano (CEB)} & \multicolumn{2}{c|}{\bf Equalized }  & 
\multicolumn{2}{c|}{\bf Unequalized }  
\\ 
\cline{2-13}
  &\multicolumn{2}{c|}{Male} & \multicolumn{2}{c|}{Female} & 
\multicolumn{2}{c|}{Male} & \multicolumn{2}{c|}{Female} & 
\multicolumn{2}{c|}{Male+Female} &
\multicolumn{2}{c|}{Male+Female} \\
\cline{2-13}
 & EER(\%) & minC &  EER(\%) & minC  &  EER(\%) & minC  & 
 EER(\%) & minC &  EER(\%) & minC  &  EER(\%) & minC  \\ \hline\hline
Fusion A &
5.95 & 0.235 & 15.24 & 0.546 & 22.77 & 0.841 & 22.43 & 0.788 & 16.60 & 0.602 & 16.71 & 0.688 \\ 
\hline
Fusion B &
5.39 & 0.224 & 15.64 & 0.574 & 22.83 & 0.845 & 22.39 & 0.794 & 16.56 & 0.609 & 16.90 & 0.669 \\ 
\hline
\end{tabular}
\caption{\label{table:fusion}Performance of fusion by mean of 
scores from single systems 1 to 4 (Fusion A) and 1 to 8 (Fusion B).}
\end{table*}

\subsection{Post-normalization}

In \cite{Ioffe06}, a post-PLDA normalization procedure is proposed that
simultaneously diagonalizes between- and within-class covariance provided by
the PLDA learning described above. Given $\mathbf{B}$ and $\mathbf{W}$
matrices estimated by PLDA learning, the following transformation is applied:

\begin{itemize}
\item Transform data: 
$\mathbf{w}\rightarrow\mathbf{W}^{-\frac{1}{2}}\mathbf{w}$ (or, equivalently,
$\mathbf{w}\rightarrow\mathbf{L}^{-1}\mathbf{w}$ where $\mathbf{W=LL}^{t}$ is
the Cholesky decomposition of $\mathbf{W}$),

\item Compute SVD of $\mathbf{W}^{-\frac{1}{2}}\mathbf{BW}^{-\frac{1}{2}%
}=\mathbf{P\Psi P}^{t}$, where $\mathbf{P}$ is the eigenvector matrix and
$\mathbf{\Psi}$ is the diagonal matrix of eigenvalues,

\item Rotate data: $\mathbf{w}\rightarrow\mathbf{P}^{t}\mathbf{w}$.
\end{itemize}

By this way, matrices $\mathbf{B}$ and $\mathbf{W}$ become diagonal matrices
$\mathbf{\Psi}$ and $\mathbf{I}$, where $\mathbf{I}$ is the identity matrix.

As observed in \cite{Garcia11}, we assume that length-normalizing the test
data after this procedure will contribute to further reduce the shift between
training and test data. Thus, our post-normalization procedure turns out to be
equivalent to the $\mathbf{LW}$-normalization described above.

It can be noticed that, after this post-normalization which diagonalizes
covariance matrices, it is shown in \cite{Ioffe06} that reducing the
dimensionality of the between-class variability (\textit{eigenvoice} subspace
rank) can be easily achieved by keeping the largest elements of $\mathbf{\Psi
}$ and setting the rest to zero. Thus, this transformation allows fast
estimation of PLDA parameters with various eigenvoice ranks.

\section{LIA submission results}
\label{sec:results}

Total 8 sub-systems are constructed by various front-end and back-end combinations as summarized in Table~\ref{table:performance}. The Equal Error Rate (EER) and minC$_{primary}$ (minC) cost functions obtained from these systems are shown, also detailed performance by gender and language (Cebuano and Mandarin).

We note that the principal metric is minC$_{primary}$ and therefore the systems and sub-systems are optimized on this metric.


\noindent Table~\ref{table:fusion} shows the performance gain obtained by fusion of single system scores. ``Fusion A'' is the score obtained by the mean of scores (equal weights) from single systems 1 to 4 presented in Table \ref{table:performance}. We note that the system is the primary system of NIST SRE'16. ``Fusion B'' is the score obtained by the mean  of scores from the 8 single systems. We note that the system is one of the secondary systems of NIST SRE'16.

\section{Computation Time and Memory}
\label{sec:computationtime}

Table~\ref{tbl:cpu-time} shows the CPU time and memory requirements for computing the score of one verification trial (for the sub-system 2). All tasks were implemented in C++.

\begin{table*}
\centering
\caption{\label{tbl:cpu-time}Computation time and memory consumption of various part of the system to produce the score of one verification trial. All tasks were performed on a 64-bit Linux server with 512G RAM and an AMD Opteron Processor 6238 running at 2.6GHz. All CPU times are based on one core of the processor.}
\begin{tabular}{|l|l||c|c|}
\hline
 {Task} & Task Name & {CPU Time (sec.) per Utt.} & {Memory consumption (MB)} \\ \hline\hline
 1 & MFCC Extraction                 & 0.91       & 3.6  \\ 
 2 & Voice Activity Detection        & 0.75      & 7.3  \\ 
 3 & Computing statistics             & 1.33     & 810  \\ 
 4 & I-vector Extraction             & 0.66     & 810  \\ 
 5 & PLDA Scoring      & 0.01       &  54  \\ \hline\hline
 & Overall                         & 3.66                    &  --    \\ \hline
\end{tabular}
\end{table*}

\section{Conclusion}
We have described the LIA site speaker recognition system submitted to the 2016 NIST SRE. The systems developed were a fusion of i-vector based sub-systems using different front-end and back-end.

\newpage

\bibliographystyle{IEEEbib}
\bibliography{references}
\end{document}